\documentclass[twocolumn,secnumarabic,amssymb, nobibnotes, aps, prd, superscriptaddress]{revtex4-1}

\usepackage[T1]{fontenc}
\usepackage{graphicx}
\usepackage{color}
\usepackage{amsmath}
\usepackage{upgreek}
\usepackage{gensymb}
\usepackage[left,modulo]{lineno}
\usepackage{xcolor}

\usepackage{hyperref}

\AtBeginDocument{\nocite{achemso-control}}

\begin{document}
\title{Viscoelasticity induced onset of slip at the wall for polymer fluids}

\author{Marion Grzelka}
\affiliation{Universit\'e Paris-Saclay, CNRS, Laboratoire de Physique des Solides, 91405, Orsay, France}
\author{Iurii Antoniuk}
\affiliation{Univ Lyon, Universit\'e Lyon 1, CNRS, Ing\'enierie des Mat\'eriaux Polym\`eres, UMR 5223, F-69003, Lyon, France}
\author{Eric Drockenmuller}
\affiliation{Univ Lyon, Universit\'e Lyon 1, CNRS, Ing\'enierie des Mat\'eriaux Polym\`eres, UMR 5223, F-69003, Lyon, France}
\author{Alexis Chennevi\`ere}
\affiliation{Laboratoire L\'eon Brillouin, CEA Saclay, 91191 Gif-sur-Yvette, France}
\author{Liliane L\'eger}
\affiliation{Universit\'e Paris-Saclay, CNRS, Laboratoire de Physique des Solides, 91405, Orsay, France}
\author{Fr\'ed\'eric Restagno}
\email[Corresponding author: ]{frederic.restagno@u-psud.fr}
\affiliation{Universit\'e Paris-Saclay, CNRS, Laboratoire de Physique des Solides, 91405, Orsay, France}
\date{\today}

\begin{abstract}

 The progressive onset of slip at the wall, which corresponds to a slip length  increasing with the solicitation time before reaching a plateau, has been investigated for model viscoelastic polymer solutions, allowing one to vary the longest relaxation time while keeping constant solid - fluid interactions. A hydrodynamic model based on a Maxwell fluid and the classical Navier's hypothesis of a linear response for the friction stress at the interface fully accounts for the data. In the limit of the linear viscoelasticity of the fluid, we could postulate a Newtonian response for the interfacial friction coefficient reflecting the local character of solid-liquid friction mechanisms. Deviations between the experiments and our model are observed when the fluid is far from its linear viscoelastic behavior.\\

\doi{10.1021/acsmacrolett.0c00182}
\end{abstract}
\maketitle

The description of liquid flows close to a surface has become a major issue during the last 10 years, due to the fast development of micro and nanofluidics, and driven by potential applications, as for example desalination \cite{holt_fast_2006,majumder_mass_2011,qin_measurement_2011,baek_high_2014,radha_molecular_2016,abraham_tunable_2017} or blue energy \cite{ajdari_giant_2006,pennathur_energy_2007,ren_slip-enhanced_2008,bouzigues_nanofluidics_2008,goswami_energy_2010}. 

In any fluid mechanics problem, two equations are needed: a bulk constitutive equation to describe the fluid properties, and an interfacial constitutive equation to describe the interfacial friction. For simple fluids, a single viscosity $\eta$, can be defined, assuming that the friction between two layers of fluid sliding past each other is proportional to their difference in velocity: $\sigma_\mathrm{visc}=\eta\dot{\gamma}$, where $\sigma_\mathrm{visc}$ is the viscous stress and $\dot{\gamma}$ the shear rate of the flow. Combined with Newton's laws this leads to the classical  Navier-Stokes equation. To solve this bulk differential equation, a boundary condition needs be specified. The no slip boundary condition (equal solid and fluid velocities), has long been commonly used. In the past decades however, numerous violations of this no slip boundary condition have been reported, both experimentally \cite{pit_direct_2000,zhu_rate-dependent_2001,cottin-bizonne_boundary_2005,secchi_massive_2016,cottin-bizonne_nanohydrodynamics_2008, durliat_influence_1997,bonaccurso_surface_2003,boukany_molecular_2010,cuenca_submicron_2013,baumchen_slip_2009,ilton_adsorption-induced_2018, cross_wall_2018,boukany_interfacial_2006,henot_friction_2018} and numerically \cite{thompson_shear_1990,bocquet_hydrodynamic_1994,thompson_general_1997,priezjev_molecular_2004,chen_determining_2015,omori_full_2019}, especially when polymers were involved.

 Navier \cite{navier_memoire_1823} was the first to introduce the possibility of slip at the wall, assuming a linear constitutive equation at the interface: the friction stress of a layer of fluid sliding over the solid surface, $\sigma_\mathrm{friction}$ is proportional to the slip velocity $V$: $\sigma_\mathrm{friction}=kV$, where $k$ is the so-called Navier's or friction coefficient. $k$ quantifies the ability of a given fluid to slip on a given solid surface. Balancing friction  and viscous stresses, one can define the distance to the interface where the velocity profile extrapolates to zero, the so-called slip length $b$: $b=\eta/k$. The order of magnitude of the slip length varies from few nanometers \cite{pit_direct_2000,cottin-bizonne_boundary_2005,secchi_massive_2016,cottin-bizonne_nanohydrodynamics_2008,bonaccurso_surface_2003,thompson_shear_1990,bocquet_hydrodynamic_1994,thompson_general_1997,chen_determining_2015} for simple fluids to micrometers \cite{zhu_rate-dependent_2001,durliat_influence_1997,boukany_molecular_2010,cuenca_submicron_2013,baumchen_slip_2009,ilton_adsorption-induced_2018,cross_wall_2018,boukany_interfacial_2006,henot_friction_2018} for complex fluids such as entangled polymer solutions or melts.
 It has also been demonstrated that for soft elastic solids sliding on a solid surface, the friction stress was also obeying a Navier's equation, with a friction stress proportional to the sliding velocity, and on non adsorbing surfaces, the obtained Navier's friction coefficient was exactly the same for a crosslinked elastomer and for a sheared polymer melt of the same chemical nature, except for the crosslinks \cite{henot_friction_2018}.
 This in turn rises a new question: when viscoelastic fluids are involved, is it sufficient to postulate a single constant friction coefficient, or, as for bulk fluid, is it necessary to introduce a viscoelastic response for the interfacial stress? 
Indeed, when simple or complex fluids are solicited at times smaller than their longest characteristic time, they can no longer be considered as Newtonian and the bulk fluid response presents an elastic contribution at short times, with characteristic times which strongly depend on the nature of the molecular interactions, and is in the nanosecond range for simple fluids while it can become much longer than seconds for complex fluids. The shear stress then depends on the shear time and the bulk constitutive equation has to be modified. 
Is it necessary to introduce a time dependent Navier coefficient, going from a solid friction coefficient at short times to the classical Navier coefficient at longer times? If so, what physico-chemical phenomena rule the characteristic time of this change in friction response of the interface? Such a time dependent Navier coefficient has  rarely been reported \cite{thompson_general_1997,zhu_rate-dependent_2001,craig_shear-dependent_2001,priezjev_molecular_2004,cross_wall_2018,omori_full_2019} in the literature and an universal interfacial constitutive equation is still lacking to pave the way to what could become solid-liquid interfacial rheology.

Inspired by model bulk rheology experiments, we present an investigation the interfacial friction for series of model complex fluid in contact with two different model surfaces, and submitted to an abrupt change of shear. To do so, we used entangled polymer solutions, for which the longest characteristic time, the terminal time, or reptation time $\tau_\mathrm{rep}$, can easily be adjusted over a large range through the polymer volume fraction $\phi$, while keeping constant the local fluid - solid interactions. For such fluids, the slip length, directly linked to the interfacial stress, ranges from tens to thousands of micrometers, large enough to be easily measured. We put into evidence a progressive onset of slip at the wall, characterized by a dependence of the slip length versus the shear time, before reaching a steady state slip regime. All characteristics of this transient slip regime are deeply affected by the polymer volume fraction. To rationalize these results, we built a model for a Maxwell-like fluid in simple shear flow, and obeying a classical Navier's time independent boundary condition. The comparison between experimental data for the time evolution of the slip length during the onset of slip and this model allows one to draw some conclusions on the locality of the fluid - solid friction mechanisms. 

\textbf{Experimental approach.}

\begin{figure}[htbp]
  \centering
  \includegraphics[width=7.5cm]{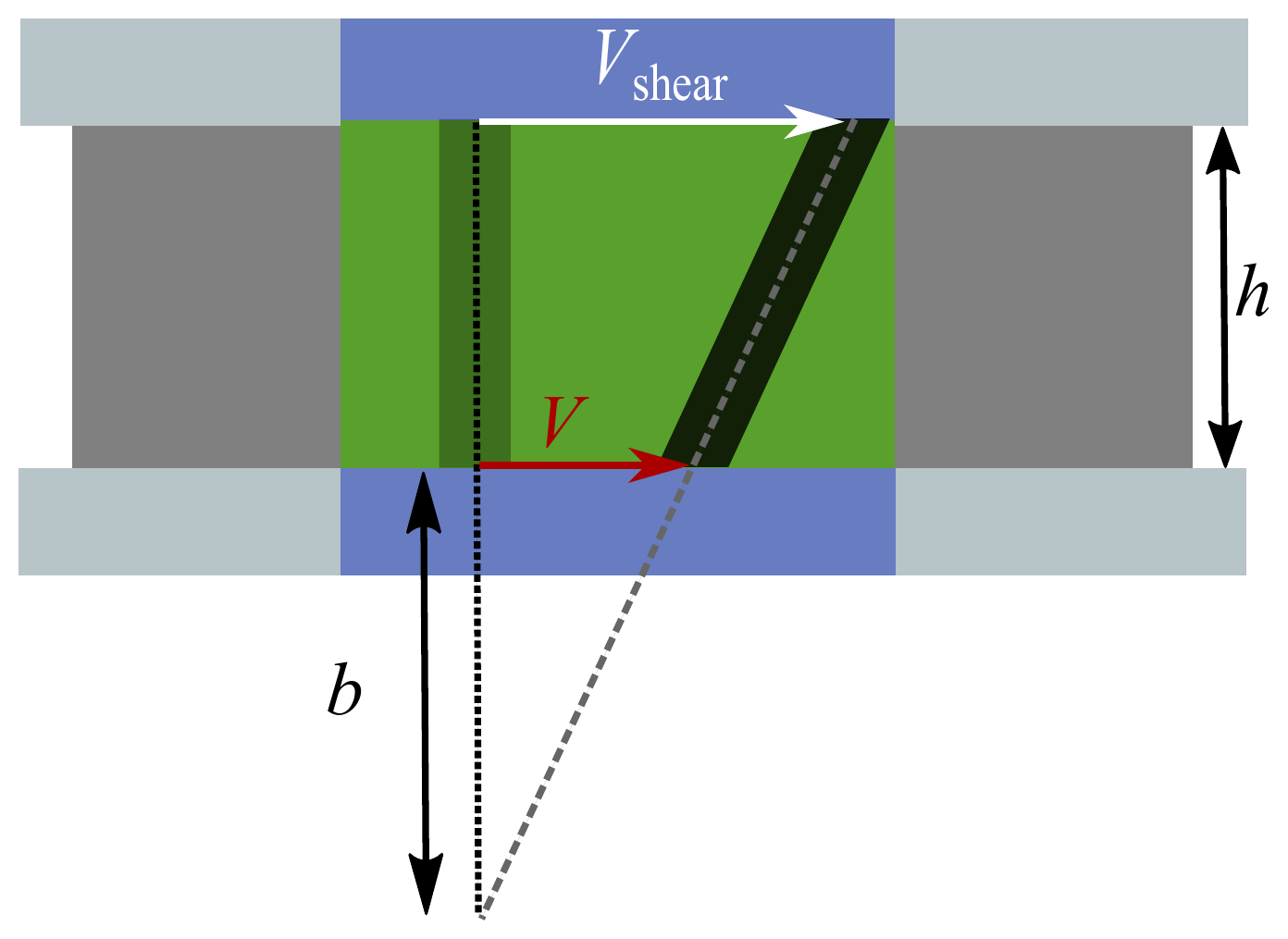}
  \caption{Measurement of the slip length $b$ of fluids by velocimetry using photobleaching. A drop of fluid, of thickness $h$, is sheared at a constant velocity $V_\mathrm{shear}$ during a time $t$. The fluid slips with a velocity $V$ on the bottom surface.}
  \label{1}
\end{figure}

Velocimetry using fluorescence photobleaching is commonly used to measure slip lengths in polymer fluids (Fig.\ref{1}). A drop of fluorescent photobleachable fluid  is compressed between two plane solid surfaces. The thickness $h$ of the drop is measured by spectroscopic reflectometry. A pattern is photobleached in the fluid which is then sheared during a monitored time $t$ by displacing the top surface at a constant velocity $V_\mathrm{shear}=d_\mathrm{shear}/t$, where $d_\mathrm{shear}$ is the total displacement of the top surface. The slip length at the bottom surface is directly measured by following the evolution of the displaced photobleached pattern, as previously described \cite{henot_friction_2018}.

To adjust the characteristic times of the studied viscoelastic fluids, semi-dilute solutions with tunable volume fraction $\phi$ of polystyrene (PS, $M_\mathrm{n} = 10.2\,\mathrm{Mg}\cdot\mathrm{mol}^{-1}$, \textit{\DJ} = 1.08, Polymer Source Inc.) in diethyl phthalate (DEP) were used. High molar mass PS and ca. $1\mathrm{wt}\%$  of a photobleachable
polystyrene ($M_\mathrm{n} = 429 \,\mathrm{kg}\cdot\mathrm{mol}^{-1}$, \textit{\DJ} = 1.05, see Fig. S2) were dissolved in diethyl phthalate and toluene and were gently stirred for at least 3 weeks. Toluene was then evaporated at room temperature under vacuum during a week. The fluorescently labeled polystyrene contains nitro-benzoxadiazole (NBD) fluorescent groups emitting at $550\,\mathrm{nm}$ when excited at $458\,\mathrm{nm}$ at both chain ends. Synthesis protocol and characterization of PS di-NBD are detailed in Supplementary Materials. The respective viscosity $\eta$ and reptation time $\tau_\mathrm{rep}$ were measured for each solution by oscillatory rheology at $22^\circ\,\mathrm{C}$ using an Anton-Paar MCR 302 rheometer in a cone plate geometry ($2^\circ$ cone angle, $25\,\mathrm{mm}$ diameter). A classical viscoelastic behavior is evidence for all polymer solutions (see Fig. S4 ). 

Table \ref{tab6:solutions_PS} summarizes the measured characteristics of the PS in DEP solutions.






\begin{table}[htbp]
\centering
\begin{tabular}{c c c c c}
\hline\hline
     $\phi$ & $\tau_\mathrm{rep}$ [s] &$ \eta$ [Pa$\cdot$s] & $ b_{\mathrm{\infty,\,PS}}$ [$\upmu$m] & $b_{\mathrm{\infty,Si}}$ [$\upmu$m] \\
\hline
    0.0230 & 2.7 &  64  & $44\pm 1$ & $58\pm 3$\\

    0.0314 & 8.3 & 401 & $99 \pm 2$& $184\pm 5$\\

    0.0397 & 16.3 & 1,147 & $249\pm 6$ & $327\pm 7$\\

    0.0495 & 24.3 & 3,840 & $628\pm 13$ & $1,254\pm 37$\\

    0.0608 & 50 & 12,000 &  $1,217 \pm 37$& $2,116\pm 31$\\
\hline\hline
\end{tabular}
\caption{Experimental characteristics of semi-dilute solutions of PS in DEP.}
\label{tab6:solutions_PS}
\end{table}

Slip lengths for 5 solutions with $\phi$ ranging from $0.023$ to $0.061$ were measured on two model substrates: a bare silicon wafer and a dense layer of grafted-to PS brushes. Bare silicon wafer (2" diameter, $3\,\mathrm{mm}$ width, Si-Mat Inc.) was cleaned before each slip measurement with a UV/O$_3$ treatment for $30\,\mathrm{min}$. The layer of dense PS brushes was prepared following a previously published protocol \cite{chenneviere_direct_2016}. Briefly, a self-assembled monolayer of triethoxy(3-glycidyloxypropyl)-silane was vapor deposited on a Si wafer. The SAM was $0.9\,\mathrm{nm}$ thick as measured by ellipsometry. Amino end-functionalized PS ($M_\mathrm{n}=5.0\,\mathrm{kg}\cdot\mathrm{mol}^{-1}$, \textit{\DJ} $=1.17$, Polymer Source Inc.) was covalently tethered to the SAM in the melt at $140\,^\circ$C for 48 hours. After through rinsing of the non-covalently tethered chains, the grafted PS layer was $2.8\,\mathrm{nm}$ thick and was considered as a dense polymer brush. The solutions are sheared at various shear rates in the range $[2\times10^{-4}-5.7]\,\mathrm{s}^{-1}$.

\begin{figure}[htbp]
  \centering
     \includegraphics[width=\columnwidth]{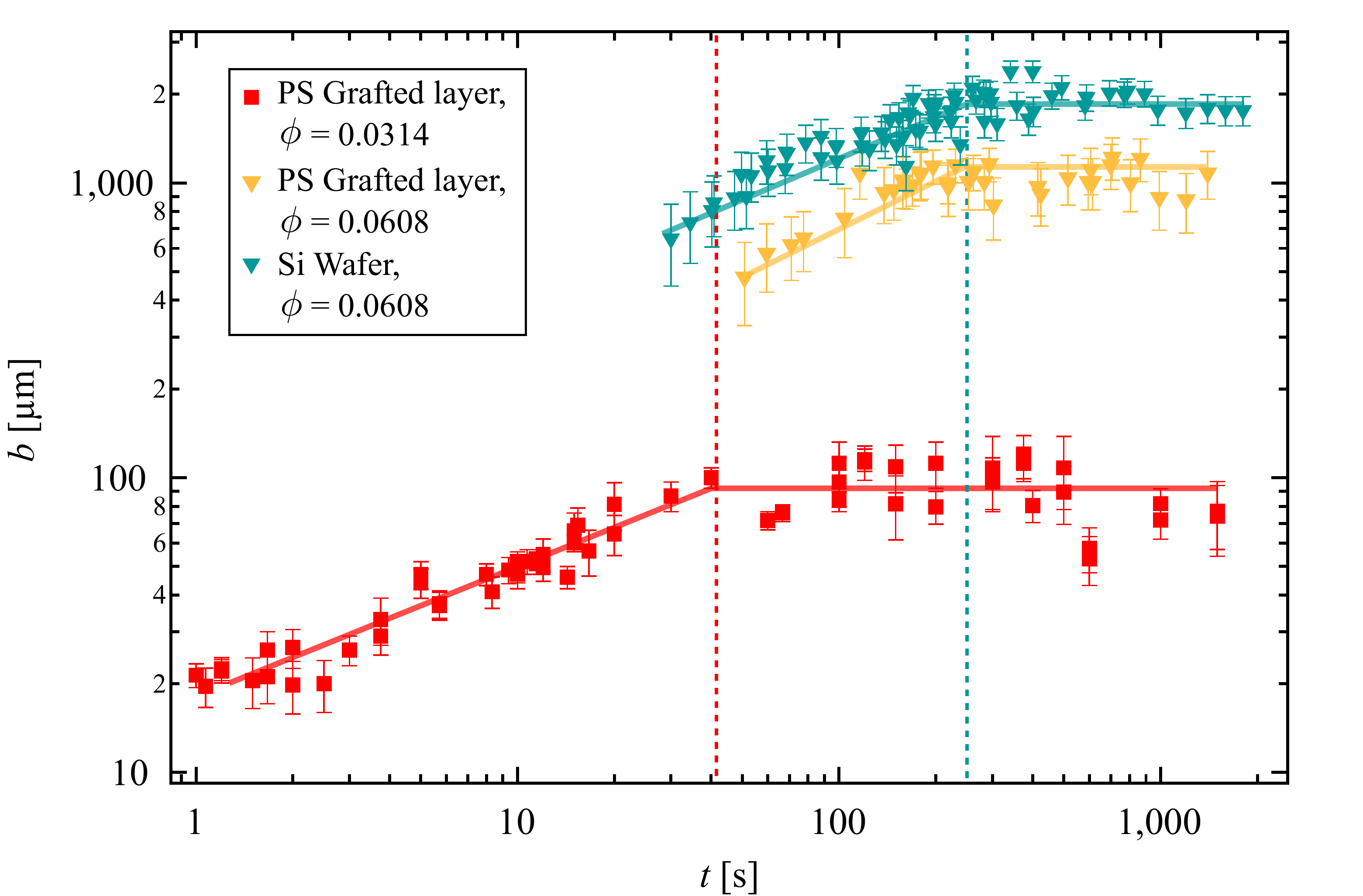}
  \caption{Transient onset of slippage for polymer solutions. Slip length $b$ of PS in DEP solutions as a function of the shear time $t$ on a bare silicon wafer and on a grafted layer of PS brushes, for different PS volume fractions $\phi$.The dotted lines represent $t=5\tau_\mathrm{rep}$, a typical time for which the steady state regime is reached. The solid lines are guides to the eye.}

  \label{2}
\end{figure}

As shown in Figure \ref{2} for 3 representative systems, the slip length $b$ increases with the shear time $t$ until it reaches a plateau. Similar results have been observed for all 5 volume fractions on the two substrates (see Supplementary Fig.S3). These data clearly evidence a transient behavior, corresponding to a progressive onset of slip before reaching a steady state regime. The characteristic time of the transient regime (dotted lines in Fig.\ref{2}) depends on $\phi$ but not on the substrate. Conversely, the plateau value obtained in the steady state regime, noted $b_\infty$, strongly depends on both $\phi$ and the substrate. This complex dependence is interesting by itself and highlights that the stress transmission at the interface depends on both the local structure of the interface and on the bulk properties \cite{mhetar_slip_1998,plucktaveesak_interfacial_1999,sanchez-reyes_interfacial_2003}. The discussion of the influence of the concentration on the slip length will be developed in a forthcoming paper. The transient behavior appears to be independent of the shear velocity $V_\mathrm{shear}$, as the measurements presented in Figure \ref{2} have been obtained for different values of $V_\mathrm{shear}$.

In order to gain a better understanding of the observed transient regime, we built a simple mechanical model, based on the Navier's boundary condition and Maxwell-like fluids.

The polymer solution is considered as a Maxwell-like fluid, characterized by its viscosity $\eta$ in the linear regime and its elastic modulus $E$, directly linked to its reptation time $\tau_\mathrm{rep}$: $\tau_\mathrm{rep}=\eta/E$. 
As the slip length $b$ may depend on the shear time $t$, the real shear rate experienced by the fluid also depends on $t$, even for a constant shear velocity $V_\mathrm{shear}=d_\mathrm{shear}/t$. In the simple shear geometry illustrated in Fig. \ref{1}, one can link the shear rate dependence with $t$ to $b(t)$:
\begin{align}
    \dot{\gamma}(t)&=\frac{V_\mathrm{shear}-V(t)}{h}\label{eq6:gam_dot_var_V}\\
                &=\frac{V_\mathrm{shear}[h+b(t)]-V_\mathrm{shear}t\dot{b}(t)}{[h+b(t)]^2}
\label{eq6:gam_dot_var_b}
\end{align}

In a Maxwell-like fluid, one can write the viscous stress $\sigma_\mathrm{visc}(t)=\eta\dot{\gamma}_\mathrm{visc}(t)$ and the elastic stress $\sigma_\mathrm{elas}(t)=E\gamma_\mathrm{elas}(t)$. At the solid/liquid interface, assuming the friction coefficient  to be independent of the shear time, Navier's hypothesis gives $\sigma_\mathrm{friction}=kV(t)$. The stress balance at the interface and the expression of the imposed shear strain $\gamma_\mathrm{tot}=\gamma_\mathrm{visc}+\gamma_\mathrm{elas}$ lead to a differential equation for the shear stress. Considering a zero shear stress at the beginning of the shear, the unique solution to this differential equation writes:

\begin{equation}
    \sigma(t)=\frac{V_\mathrm{shear}E\tau}{h}\left(1-e^{-t/\tau}\right)
    \label{eq6:sigma(t)}
\end{equation}
where $\tau=\frac{\tau_\mathrm{rep}}{1+\frac{b_\mathrm{\infty}}{h}}$ is a characteristic time depending on the steady-state slippage through $b_\mathrm{\infty}=\frac{\eta}{k}$. Equation \eqref{eq6:sigma(t)} is the expected expression of the shear stress for a Maxwell-like fluid, except for the fact that the characteristic time $\tau$ now depends on the amount of slip at the wall.The total shear rate $\dot{\gamma}$ may also be calculated and the viscous and elastic parts evaluated separately (cf SI). 

Pearson and Petrie \cite{pearson_melt-flow_1968} were the first to introduce theoretically the notion of "retarded slip" due to the presence of a relaxation slip time $\lambda_s$, with a model based on an analogy with a mechanical model for Maxwell-like fluids. Considering a power-law slip model, Hill \textit{et al.} \cite{hill_apparent_1990} and Hatzikiriakos \textit{et al.} \cite{hatzikiriakos_wall_1991} postulated a differential equation for the shear velocity: $V+\lambda_s\frac{\mathrm{d}V}{\mathrm{d}t}=a\sigma^m$, where $a$ is a slip coefficient and $m$ is the slip power-law exponent. The origin of the introduced relaxation slip time $\lambda_s$ was not discussed in these papers. With our model, the postulated relaxation slip time $\lambda_s$ can be clearly identified as the differential equation obtained for $\sigma$ may be written for the slip velocity $V$ : $V+\dot{V}\tau=\frac{V_\mathrm{shear}E\tau}{kh}$. Here the characteristic time $\tau=\lambda_s$ is directly linked to the slip properties through $b_\infty$, the fluid dynamics through $\tau_\mathrm{rep}$ and  the geometry of the experiment through $h$.

Combining equation \eqref{eq6:sigma(t)} to the dependence of the shear rate with $t$ (eq. \eqref{eq6:gam_dot_var_V} and \eqref{eq6:gam_dot_var_b}) allows one to solve the differential equation obtained for $b(t)$, with the initial no slip condition: 
\begin{equation}
    b(T)=h\left[\frac{T(1+X)}{T-\frac{X}{1+X}e^{-T(1+X)}+\frac{X}{1+X}}-1 \right]
    \label{eq6:b(T)}
\end{equation}
where $T=\frac{t}{\tau_\mathrm{rep}}$ compares the shear time $t$ and the reptation time $\tau_\mathrm{rep}$. $X=\frac{b_\mathrm{\infty}}{h}$ is the ratio between the steady state slip length $b_\mathrm{\infty}$ and the thickness $h$ of the sheared drop of fluid. Taking into account Navier's hypothesis for a Maxwell-like fluid affords a complex temporal dependence of the slip length. It is worth noting that equation \eqref{eq6:b(T)} is independent of the shear velocity $V_\mathrm{shear}$ and thus of the apparent shear rate $\dot{\gamma}_\mathrm{app}=V_\mathrm{shear}/h$, in agreement with our experimental results.

\begin{figure}[htbp]
  \centering
  \includegraphics[width=\columnwidth]{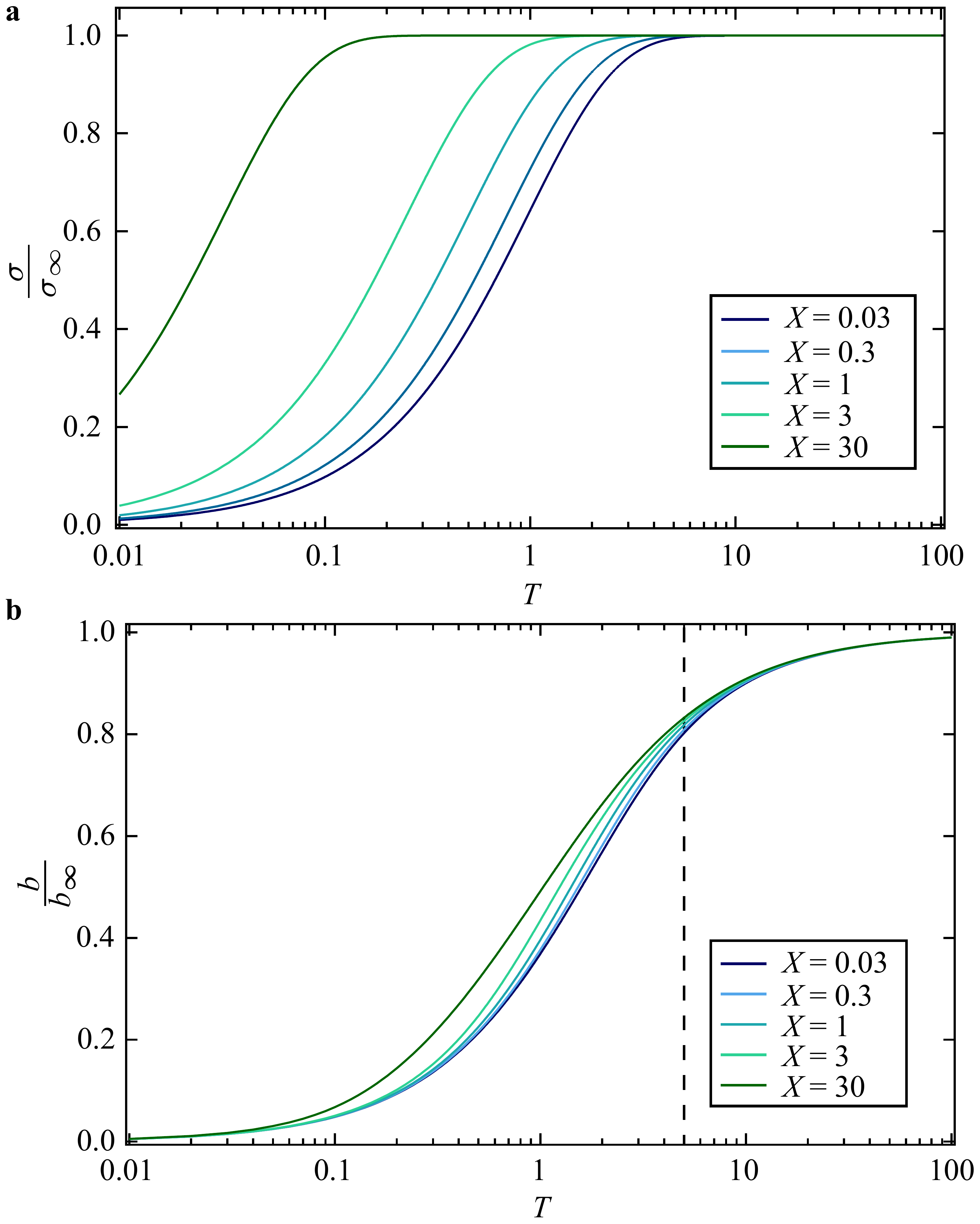}
 \caption{Slippage of a Maxwell-like fluid following the Navier's boundary condition. \textbf{a} Normalized shear stress $\sigma/\sigma_\infty$ as a function of the normalized shear time $T=t/\tau_\mathrm{rep}$.  \textbf{b} Normalized slip length $b/b_\infty$ as a function of the normalized shear time $T=t/\tau_\mathrm{rep}$. The dotted line represents $T=5$, at which the steady state regime is typically reached. All the calculations are made for 5 non-denationalized values $X=b_\infty/h$, with $h$ the height of the sheared fluid.}
  \label{3}
\end{figure}

As shown in Figure \ref{3}a, no matter what is the value of the parameter $X$, the shear stress calculated from eq. \eqref{eq6:sigma(t)} increases with the shear time, before reaching a plateau. The value of $X$ clearly affects the dynamics of the onset of slippage: the larger the thickness $h$ compared to $b_\infty$ (smaller $X$), the longer the transient slip regime. The dependence of the calculated slip length versus the shear time, presented in Figure \ref{3}b, appears however only weakly affected by the value of $X$. 

The theoretical evolution of $b(t)/b_\infty$ shown in Fig\ref{3}b, is qualitatively similar to the trend of the  experimental data: $b(t)$ increases before reaching a steady state regime. A shear time longer than few reptation times is needed to obtain an almost constant slip length. For all $X$ values, the slip length is larger than $0.8 b_\mathrm{\infty}$ at $T=5$, as indicated by the dotted line in Fig\ref{3}b.

\begin{figure}[htbp]
  \centering
   \includegraphics[width=\columnwidth]{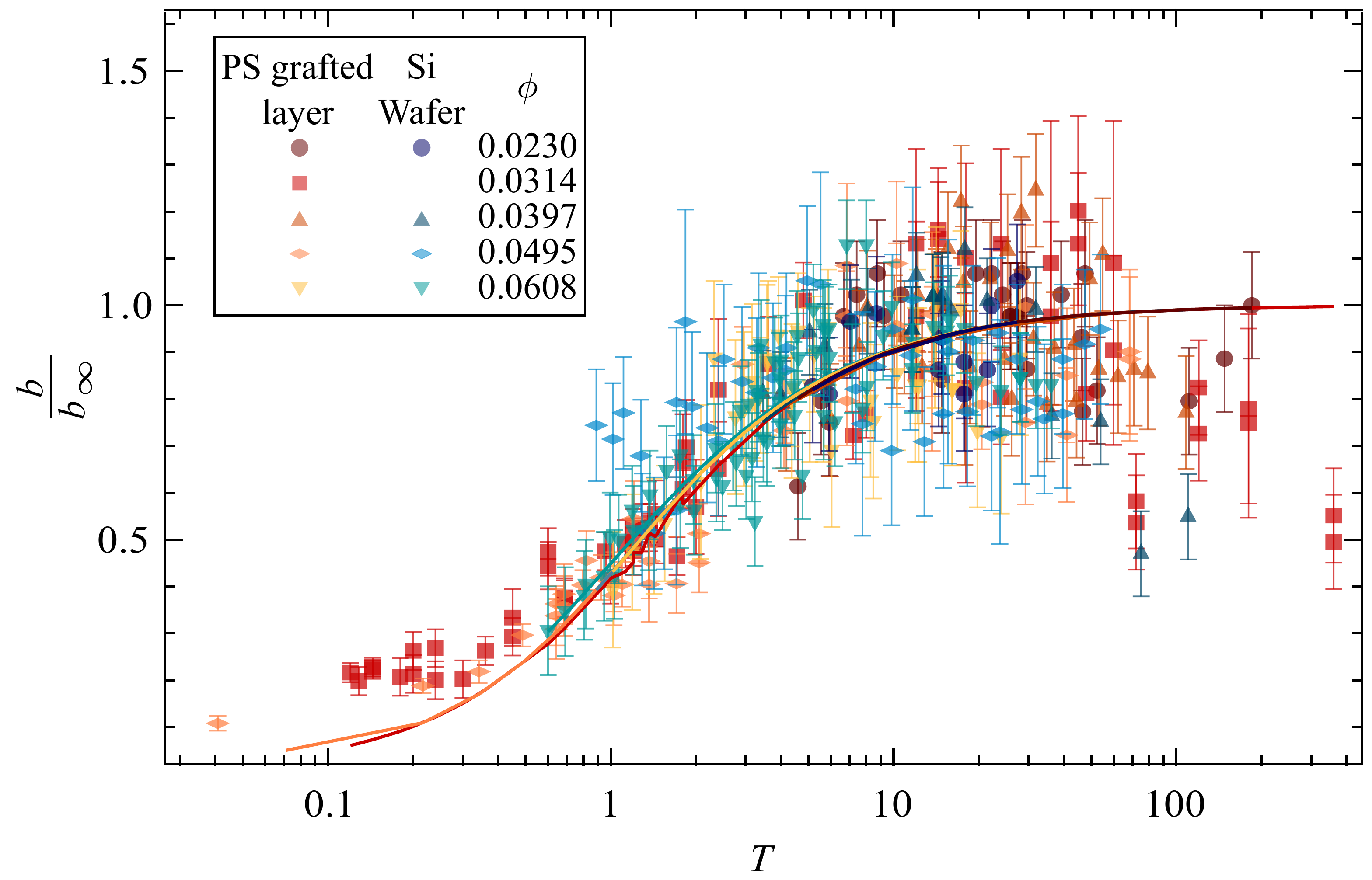}
 \caption{Comparison between measured slippage of polymer fluids and calculated Maxwell-like fluid slippage. Normalized slip length $b/b_\infty$ of PS in DEP solutions as a function of the normalized shear time $T=t/\tau_\mathrm{rep}$ on a bare silicon wafer and on a grafted layer of PS brushes. The lines correspond to the calculation by equation \eqref{eq6:b(T)}, $b_\infty$ being the only adjustable parameters.}
  \label{4}
\end{figure}

To compare quantitatively the experimental data to the prediction of eq. \eqref{eq6:b(T)}, we used the following procedure. The thickness $h$ of the drop, the reptation time $\tau_\mathrm{rep}$ and the slip length depending on the shear time $b(t)$ are measured independently. We fitted the experimental data with eq.\eqref{eq6:b(T)} with $b_\mathrm{\infty}$ as an adjustable parameter. Table \ref{tab6:solutions_PS} summarizes the fitted values for $b_\mathrm{\infty}$.
Figure \ref{4} presents the measured (symbols) and calculated (lines) normalized slip lengths $b/b_\infty$ as a function of the normalized shear time $T=t/\tau_\mathrm{rep}$. All experimental data  (five solutions and two surfaces) collapse on a master curve, indicating a universal behavior, only depending on both the volume fraction $\phi$ and the shear time $t$, but independent of the solid substrate. 
Note that in the experiments, the thickness $h$ of the sheared drop is chosen to be as close as possible to $b_\infty$ in order to minimize the error bars on the measured $b$ values. Therefore experimental $X$ values only vary within one decade ($X\in[0.3 -3.8]$). The experimental values can thus be compared with a theoretical model calculated with the corresponding $X$ values.
The model clearly well captures the temporal evolution of $b(t)$.

A slight difference between model and experimental data appears for $T<1$. The experimental slip lengths estimated by our model seem underestimated compared to the experiments. These points correspond to large shear velocities. For these shear velocities, we enter in the shear-thinning regime of the fluid. This shear thinning regime is  out of the scope of the present mechanical model, which only deals with linear viscoelastic fluids.At large $T$, the time of diffusion of labeled polymers becomes comparable to the shear time, leading to underestimate the error bars for $b(t)$. The good agreement between model and experimental data in the Newtonian regime confirms that viscoelasticity combined to a linear friction fully describe the onset of slippage in these complex fluids. 

No viscoelastic response of friction at the wall is needed to account for the observed dependence of the slip length on the shearing time. This confirms the idea that the interfacial friction is a local quantity, as already  observed for polymer melts \cite{de_gennes_ecoulements_1979,baumchen_reduced_2009,henot_friction_2018}. If indeed governed by phenomena at monomer or solvent molecular scales, the characteristic relaxation times associated to interfacial friction should be several orders of magnitude smaller than the relaxation times of the model viscoelastic fluids used here, and not accessible to the present experiments. The picture of a locally determined interfacial friction could be  affected by chains adsorbed at the solid surface. The substrates used here have been specifically chosen to minimize adsorption: PS is known to adsorb slowly (several hours) on bare silicon wafers \cite{housmans_kinetics_2014}, and we have shown (see SI) that indeed a weakly dense adsorbed layer could form on silica in conditions comparable to our experiments on bare silica, while the dense brush of short PS chains prevents adsorption of high molar mass PS chains on the other investigated surface. However, we see no evidence of any effect of the evolution of such an adsorbed layer during the slippage experiments: they were all done at random order for $V_\mathrm{shear}$ and $\phi$. We could not notice any effect of the time of contact of the solution with the surface on the obtained results. This is why the picture of a progressive onset of slippage controlled by the viscoelasticity of the fluid, while keeping a time independent Navier's coefficient appears fully efficient to account for experimental data. 

To conclude, we have provided unambiguous experimental evidences of the existence of a progressive increase in slip length with shearing time before reaching a steady state plateau, for polymer solutions flowing on ideal surfaces. A mechanical model based on the Navier's hypothesis for Maxwell-like fluids has been built to describe this transient onset of slip, and pin point the mechanisms at stake. The good agreement between model and experiments validates the main hypothesis of the model: on ideal substrates (no adsorbed chains), the friction is driven by local phenomena and no elastic contribution to interfacial friction is needed to account for the onset of slip. The model, applicable to any linear viscoelastic fluid, provides a firm framework to identify which parameters govern the onset of slip for fluids others than polymer solutions. As the shear stress appears to be more sensitive than the slip length to details of the experimental geometry in the transient slip regime, this work highlights the importance of studying the interfacial rheology at a solid/liquid interface by analyzing the temporal evolution of the interfacial shear stress. The generality of the model may also lead to a better understanding of the molecular mechanisms responsible for the instabilities observed in extrusion process. We anticipate that extending this approach will contribute to understand fluid dynamics in different regimes, such as flow in nanoconfinement and turbulence in transient regimes.

\section*{Acknowledgments:} This  work  was  supported  by  ANR-ENCORE program (ANR-15-CE06-005). We thank M. Corpart for experimental help. F.R. thanks J.D. McGraw and C. Derail for regular discussions.
\section*{ASSOCIATED CONTENT}
Supporting Information available: Experimental methods and supporting figures.



\begin{mcitethebibliography}{44}
\providecommand*\natexlab[1]{#1}
\providecommand*\mciteSetBstSublistMode[1]{}
\providecommand*\mciteSetBstMaxWidthForm[2]{}
\providecommand*\mciteBstWouldAddEndPuncttrue
  {\def\EndOfBibitem{\unskip.}}
\providecommand*\mciteBstWouldAddEndPunctfalse
  {\let\EndOfBibitem\relax}
\providecommand*\mciteSetBstMidEndSepPunct[3]{}
\providecommand*\mciteSetBstSublistLabelBeginEnd[3]{}
\providecommand*\EndOfBibitem{}
\mciteSetBstSublistMode{f}
\mciteSetBstMaxWidthForm{subitem}{(\alph{mcitesubitemcount})}
\mciteSetBstSublistLabelBeginEnd
  {\mcitemaxwidthsubitemform\space}
  {\relax}
  {\relax}

\bibitem[Holt \latin{et~al.}(2006)Holt, Park, Wang, Stadermann, Artyukhin,
  Grigoropoulos, Noy, and Bakajin]{holt_fast_2006}
Holt,~J.~K.; Park,~H.~G.; Wang,~Y.; Stadermann,~M.; Artyukhin,~A.~B.;
  Grigoropoulos,~C.~P.; Noy,~A.; Bakajin,~O. Fast {Mass} {Transport} {Through}
  {Sub}-2-{Nanometer} {Carbon} {Nanotubes}. \emph{Science} \textbf{2006},
  \emph{312}, 1034--1037\relax
\mciteBstWouldAddEndPuncttrue
\mciteSetBstMidEndSepPunct{\mcitedefaultmidpunct}
{\mcitedefaultendpunct}{\mcitedefaultseppunct}\relax
\EndOfBibitem
\bibitem[Majumder \latin{et~al.}(2011)Majumder, Chopra, and
  Hinds]{majumder_mass_2011}
Majumder,~M.; Chopra,~N.; Hinds,~B.~J. Mass {Transport} through {Carbon}
  {Nanotube} {Membranes} in {Three} {Different} {Regimes}: {Ionic} {Diffusion}
  and {Gas} and {Liquid} {Flow}. \emph{ACS Nano} \textbf{2011}, \emph{5},
  3867--3877\relax
\mciteBstWouldAddEndPuncttrue
\mciteSetBstMidEndSepPunct{\mcitedefaultmidpunct}
{\mcitedefaultendpunct}{\mcitedefaultseppunct}\relax
\EndOfBibitem
\bibitem[Qin \latin{et~al.}(2011)Qin, Yuan, Zhao, Xie, and
  Liu]{qin_measurement_2011}
Qin,~X.; Yuan,~Q.; Zhao,~Y.; Xie,~S.; Liu,~Z. Measurement of the {Rate} of
  {Water} {Translocation} through {Carbon} {Nanotubes}. \emph{Nano Letters}
  \textbf{2011}, \emph{11}, 2173--2177\relax
\mciteBstWouldAddEndPuncttrue
\mciteSetBstMidEndSepPunct{\mcitedefaultmidpunct}
{\mcitedefaultendpunct}{\mcitedefaultseppunct}\relax
\EndOfBibitem
\bibitem[Baek \latin{et~al.}(2014)Baek, Kim, Seo, Kim, Lee, Kim, Ahn, Bae, Lee,
  Lim, Lee, and Yoon]{baek_high_2014}
Baek,~Y.; Kim,~C.; Seo,~D.~K.; Kim,~T.; Lee,~J.~S.; Kim,~Y.~H.; Ahn,~K.~H.;
  Bae,~S.~S.; Lee,~S.~C.; Lim,~J.; Lee,~K.; Yoon,~J. High performance and
  antifouling vertically aligned carbon nanotube membrane for water
  purification. \emph{Journal of Membrane Science} \textbf{2014}, \emph{460},
  171--177\relax
\mciteBstWouldAddEndPuncttrue
\mciteSetBstMidEndSepPunct{\mcitedefaultmidpunct}
{\mcitedefaultendpunct}{\mcitedefaultseppunct}\relax
\EndOfBibitem
\bibitem[Radha \latin{et~al.}(2016)Radha, Esfandiar, Wang, Rooney, Gopinadhan,
  Keerthi, Mishchenko, Janardanan, Blake, Fumagalli, Lozada-Hidalgo, Garaj,
  Haigh, Grigorieva, Wu, and Geim]{radha_molecular_2016}
Radha,~B. \latin{et~al.}  Molecular transport through capillaries made with
  atomic-scale precision. \emph{Nature} \textbf{2016}, \emph{538},
  222--225\relax
\mciteBstWouldAddEndPuncttrue
\mciteSetBstMidEndSepPunct{\mcitedefaultmidpunct}
{\mcitedefaultendpunct}{\mcitedefaultseppunct}\relax
\EndOfBibitem
\bibitem[Abraham \latin{et~al.}(2017)Abraham, Vasu, Williams, Gopinadhan, Su,
  Cherian, Dix, Prestat, Haigh, Grigorieva, Carbone, Geim, and
  Nair]{abraham_tunable_2017}
Abraham,~J.; Vasu,~K.~S.; Williams,~C.~D.; Gopinadhan,~K.; Su,~Y.;
  Cherian,~C.~T.; Dix,~J.; Prestat,~E.; Haigh,~S.~J.; Grigorieva,~I.~V.;
  Carbone,~P.; Geim,~A.~K.; Nair,~R.~R. Tunable sieving of ions using graphene
  oxide membranes. \emph{Nature Nanotechnology} \textbf{2017}, \emph{12},
  546--550\relax
\mciteBstWouldAddEndPuncttrue
\mciteSetBstMidEndSepPunct{\mcitedefaultmidpunct}
{\mcitedefaultendpunct}{\mcitedefaultseppunct}\relax
\EndOfBibitem
\bibitem[Ajdari and Bocquet(2006)Ajdari, and Bocquet]{ajdari_giant_2006}
Ajdari,~A.; Bocquet,~L. Giant {Amplification} of {Interfacially} {Driven}
  {Transport} by {Hydrodynamic} {Slip}: {Diffusio}-{Osmosis} and {Beyond}.
  \emph{Physical Review Letters} \textbf{2006}, \emph{96}, 186102\relax
\mciteBstWouldAddEndPuncttrue
\mciteSetBstMidEndSepPunct{\mcitedefaultmidpunct}
{\mcitedefaultendpunct}{\mcitedefaultseppunct}\relax
\EndOfBibitem
\bibitem[Pennathur \latin{et~al.}(2007)Pennathur, Eijkel, and Van
  Den~Berg]{pennathur_energy_2007}
Pennathur,~S.; Eijkel,~J. C.~T.; Van Den~Berg,~A. Energy conversion in
  microsystems: is there a role for micro/nanofluidics? \emph{Lab on a Chip}
  \textbf{2007}, \emph{7}, 1234--1237\relax
\mciteBstWouldAddEndPuncttrue
\mciteSetBstMidEndSepPunct{\mcitedefaultmidpunct}
{\mcitedefaultendpunct}{\mcitedefaultseppunct}\relax
\EndOfBibitem
\bibitem[Ren and Stein(2008)Ren, and Stein]{ren_slip-enhanced_2008}
Ren,~Y.; Stein,~D. Slip-enhanced electrokinetic energy conversion in
  nanofluidic channels. \emph{Nanotechnology} \textbf{2008}, \emph{19},
  195707\relax
\mciteBstWouldAddEndPuncttrue
\mciteSetBstMidEndSepPunct{\mcitedefaultmidpunct}
{\mcitedefaultendpunct}{\mcitedefaultseppunct}\relax
\EndOfBibitem
\bibitem[Bouzigues \latin{et~al.}(2008)Bouzigues, Tabeling, and
  Bocquet]{bouzigues_nanofluidics_2008}
Bouzigues,~C.~I.; Tabeling,~P.; Bocquet,~L. Nanofluidics in the {Debye} {Layer}
  at {Hydrophilic} and {Hydrophobic} {Surfaces}. \emph{Physical Review Letters}
  \textbf{2008}, \emph{101}, 114503\relax
\mciteBstWouldAddEndPuncttrue
\mciteSetBstMidEndSepPunct{\mcitedefaultmidpunct}
{\mcitedefaultendpunct}{\mcitedefaultseppunct}\relax
\EndOfBibitem
\bibitem[Goswami and Chakraborty(2010)Goswami, and
  Chakraborty]{goswami_energy_2010}
Goswami,~P.; Chakraborty,~S. Energy {Transfer} through {Streaming} {Effects} in
  {Time}-{Periodic} {Pressure}-{Driven} {Nanochannel} {Flows} with
  {Interfacial} {Slip}. \emph{Langmuir} \textbf{2010}, \emph{26},
  581--590\relax
\mciteBstWouldAddEndPuncttrue
\mciteSetBstMidEndSepPunct{\mcitedefaultmidpunct}
{\mcitedefaultendpunct}{\mcitedefaultseppunct}\relax
\EndOfBibitem
\bibitem[Pit \latin{et~al.}(2000)Pit, Hervet, and L\'eger]{pit_direct_2000}
Pit,~R.; Hervet,~H.; L\'eger,~L. Direct experimental evidence of slip in
  hexadecane: solid interfaces. \emph{Physical Review Letters} \textbf{2000},
  \emph{85}, 980--983\relax
\mciteBstWouldAddEndPuncttrue
\mciteSetBstMidEndSepPunct{\mcitedefaultmidpunct}
{\mcitedefaultendpunct}{\mcitedefaultseppunct}\relax
\EndOfBibitem
\bibitem[Zhu and Granick(2001)Zhu, and Granick]{zhu_rate-dependent_2001}
Zhu,~Y.; Granick,~S. Rate-{Dependent} {Slip} of {Newtonian} {Liquid} at
  {Smooth} {Surfaces}. \emph{Physical Review Letters} \textbf{2001}, \emph{87},
  096105\relax
\mciteBstWouldAddEndPuncttrue
\mciteSetBstMidEndSepPunct{\mcitedefaultmidpunct}
{\mcitedefaultendpunct}{\mcitedefaultseppunct}\relax
\EndOfBibitem
\bibitem[Cottin-Bizonne \latin{et~al.}(2005)Cottin-Bizonne, Cross, Steinberger,
  and Charlaix]{cottin-bizonne_boundary_2005}
Cottin-Bizonne,~C.; Cross,~B.; Steinberger,~A.; Charlaix,~E. Boundary {Slip} on
  {Smooth} {Hydrophobic} {Surfaces}: {Intrinsic} {Effects} and {Possible}
  {Artifacts}. \emph{Physical Review Letters} \textbf{2005}, \emph{94},
  056102\relax
\mciteBstWouldAddEndPuncttrue
\mciteSetBstMidEndSepPunct{\mcitedefaultmidpunct}
{\mcitedefaultendpunct}{\mcitedefaultseppunct}\relax
\EndOfBibitem
\bibitem[Secchi \latin{et~al.}(2016)Secchi, Marbach, Nigu\`es, Stein, Siria,
  and Bocquet]{secchi_massive_2016}
Secchi,~E.; Marbach,~S.; Nigu\`es,~A.; Stein,~D.; Siria,~A.; Bocquet,~L.
  Massive radius-dependent flow slippage in carbon nanotubes. \emph{Nature}
  \textbf{2016}, \emph{537}, 210--213\relax
\mciteBstWouldAddEndPuncttrue
\mciteSetBstMidEndSepPunct{\mcitedefaultmidpunct}
{\mcitedefaultendpunct}{\mcitedefaultseppunct}\relax
\EndOfBibitem
\bibitem[Cottin-Bizonne \latin{et~al.}(2008)Cottin-Bizonne, Steinberger, Cross,
  Raccurt, and Charlaix]{cottin-bizonne_nanohydrodynamics_2008}
Cottin-Bizonne,~C.; Steinberger,~A.; Cross,~B.; Raccurt,~O.; Charlaix,~E.
  Nanohydrodynamics : {The} {Intrinsic} {Flow} {Boundary} {Condition} on
  {Smooth} {Surfaces}. \emph{Langmuir} \textbf{2008}, \emph{24},
  1165--1172\relax
\mciteBstWouldAddEndPuncttrue
\mciteSetBstMidEndSepPunct{\mcitedefaultmidpunct}
{\mcitedefaultendpunct}{\mcitedefaultseppunct}\relax
\EndOfBibitem
\bibitem[Durliat \latin{et~al.}(1997)Durliat, Hervet, and
  L\'eger]{durliat_influence_1997}
Durliat,~E.; Hervet,~H.; L\'eger,~L. Influence of grafting density on wall slip
  of a polymer melt on a polymer brush. \emph{Europhysics Letters}
  \textbf{1997}, \emph{38}, 383--388\relax
\mciteBstWouldAddEndPuncttrue
\mciteSetBstMidEndSepPunct{\mcitedefaultmidpunct}
{\mcitedefaultendpunct}{\mcitedefaultseppunct}\relax
\EndOfBibitem
\bibitem[Bonaccurso \latin{et~al.}(2003)Bonaccurso, Butt, and
  Craig]{bonaccurso_surface_2003}
Bonaccurso,~E.; Butt,~H.-J.; Craig,~V. S.~J. Surface {Roughness} and
  {Hydrodynamic} {Boundary} {Slip} of a {Newtonian} {Fluid} in a {Completely}
  {Wetting} {System}. \emph{Physical Review Letters} \textbf{2003}, \emph{90},
  144501\relax
\mciteBstWouldAddEndPuncttrue
\mciteSetBstMidEndSepPunct{\mcitedefaultmidpunct}
{\mcitedefaultendpunct}{\mcitedefaultseppunct}\relax
\EndOfBibitem
\bibitem[Boukany \latin{et~al.}(2010)Boukany, Hemminger, Wang, and
  Lee]{boukany_molecular_2010}
Boukany,~P.~E.; Hemminger,~O.; Wang,~S.-Q.; Lee,~L.~J. Molecular {Imaging} of
  {Slip} in {Entangled} {DNA} {Solution}. \emph{Physical Review Letters}
  \textbf{2010}, \emph{105}, 027802\relax
\mciteBstWouldAddEndPuncttrue
\mciteSetBstMidEndSepPunct{\mcitedefaultmidpunct}
{\mcitedefaultendpunct}{\mcitedefaultseppunct}\relax
\EndOfBibitem
\bibitem[Cuenca and Bodiguel(2013)Cuenca, and Bodiguel]{cuenca_submicron_2013}
Cuenca,~A.; Bodiguel,~H. Submicron {Flow} of {Polymer} {Solutions}: {Slippage}
  {Reduction} due to {Confinement}. \emph{Physical Review Letters}
  \textbf{2013}, \emph{110}, 108304\relax
\mciteBstWouldAddEndPuncttrue
\mciteSetBstMidEndSepPunct{\mcitedefaultmidpunct}
{\mcitedefaultendpunct}{\mcitedefaultseppunct}\relax
\EndOfBibitem
\bibitem[B\"aumchen and Jacobs(2009)B\"aumchen, and Jacobs]{baumchen_slip_2009}
B\"aumchen,~O.; Jacobs,~K. Slip effects in polymer thin films. \emph{Journal of
  Physics: Condensed Matter} \textbf{2009}, \emph{22}, 033102\relax
\mciteBstWouldAddEndPuncttrue
\mciteSetBstMidEndSepPunct{\mcitedefaultmidpunct}
{\mcitedefaultendpunct}{\mcitedefaultseppunct}\relax
\EndOfBibitem
\bibitem[Ilton \latin{et~al.}(2018)Ilton, Salez, Fowler, Rivetti, Aly,
  Benzaquen, McGraw, Rapha\"el, Dalnoki-Veress, and
  B\"aumchen]{ilton_adsorption-induced_2018}
Ilton,~M.; Salez,~T.; Fowler,~P.~D.; Rivetti,~M.; Aly,~M.; Benzaquen,~M.;
  McGraw,~J.~D.; Rapha\"el,~E.; Dalnoki-Veress,~K.; B\"aumchen,~O.
  Adsorption-induced slip inhibition for polymer melts on ideal substrates.
  \emph{Nature Communications} \textbf{2018}, \emph{9}, 1--7\relax
\mciteBstWouldAddEndPuncttrue
\mciteSetBstMidEndSepPunct{\mcitedefaultmidpunct}
{\mcitedefaultendpunct}{\mcitedefaultseppunct}\relax
\EndOfBibitem
\bibitem[Cross \latin{et~al.}(2018)Cross, Barraud, Picard, L\'eger, Restagno,
  and Charlaix]{cross_wall_2018}
Cross,~B.; Barraud,~C.; Picard,~C.; L\'eger,~L.; Restagno,~F.; Charlaix,~E.
  Wall slip of complex fluids: {Interfacial} friction versus slip length.
  \emph{Physical Review Fluids} \textbf{2018}, \emph{3}, 062001\relax
\mciteBstWouldAddEndPuncttrue
\mciteSetBstMidEndSepPunct{\mcitedefaultmidpunct}
{\mcitedefaultendpunct}{\mcitedefaultseppunct}\relax
\EndOfBibitem
\bibitem[Boukany \latin{et~al.}(2006)Boukany, Tapadia, and
  Wang]{boukany_interfacial_2006}
Boukany,~P.~E.; Tapadia,~P.; Wang,~S.-Q. Interfacial stick-slip transition in
  simple shear of entangled melts. \emph{Journal of Rheology} \textbf{2006},
  \emph{50}, 641--654\relax
\mciteBstWouldAddEndPuncttrue
\mciteSetBstMidEndSepPunct{\mcitedefaultmidpunct}
{\mcitedefaultendpunct}{\mcitedefaultseppunct}\relax
\EndOfBibitem
\bibitem[H\'enot \latin{et~al.}(2018)H\'enot, Drockenmuller, L\'eger, and
  Restagno]{henot_friction_2018}
H\'enot,~M.; Drockenmuller,~E.; L\'eger,~L.; Restagno,~F. Friction of
  {Polymers}: from {PDMS} {Melts} to {PDMS} {Elastomers}. \emph{ACS Macro
  Letters} \textbf{2018}, \emph{7}, 112--115\relax
\mciteBstWouldAddEndPuncttrue
\mciteSetBstMidEndSepPunct{\mcitedefaultmidpunct}
{\mcitedefaultendpunct}{\mcitedefaultseppunct}\relax
\EndOfBibitem
\bibitem[Thompson and Robbins(1990)Thompson, and Robbins]{thompson_shear_1990}
Thompson,~P.~A.; Robbins,~M.~O. Shear flow near solids: {Epitaxial} order and
  flow boundary conditions. \emph{Physical Review A} \textbf{1990}, \emph{41},
  6830--6837\relax
\mciteBstWouldAddEndPuncttrue
\mciteSetBstMidEndSepPunct{\mcitedefaultmidpunct}
{\mcitedefaultendpunct}{\mcitedefaultseppunct}\relax
\EndOfBibitem
\bibitem[Bocquet and Barrat(1994)Bocquet, and
  Barrat]{bocquet_hydrodynamic_1994}
Bocquet,~L.; Barrat,~J.-L. Hydrodynamic boundary conditions, correlation
  functions, and {Kubo} relations for confined fluids. \emph{Physical Review E}
  \textbf{1994}, \emph{49}, 3079--3092\relax
\mciteBstWouldAddEndPuncttrue
\mciteSetBstMidEndSepPunct{\mcitedefaultmidpunct}
{\mcitedefaultendpunct}{\mcitedefaultseppunct}\relax
\EndOfBibitem
\bibitem[Thompson and Troian(1997)Thompson, and Troian]{thompson_general_1997}
Thompson,~P.~A.; Troian,~S.~M. A general boundary condition for liquid flow at
  solid surfaces. \emph{Nature} \textbf{1997}, \emph{389}, 360--362\relax
\mciteBstWouldAddEndPuncttrue
\mciteSetBstMidEndSepPunct{\mcitedefaultmidpunct}
{\mcitedefaultendpunct}{\mcitedefaultseppunct}\relax
\EndOfBibitem
\bibitem[Priezjev and Troian(2004)Priezjev, and
  Troian]{priezjev_molecular_2004}
Priezjev,~N.~V.; Troian,~S.~M. Molecular {Origin} and {Dynamic} {Behavior} of
  {Slip} in {Sheared} {Polymer} {Films}. \emph{Physical Review Letters}
  \textbf{2004}, \emph{92}, 018302\relax
\mciteBstWouldAddEndPuncttrue
\mciteSetBstMidEndSepPunct{\mcitedefaultmidpunct}
{\mcitedefaultendpunct}{\mcitedefaultseppunct}\relax
\EndOfBibitem
\bibitem[Chen \latin{et~al.}(2015)Chen, Wang, Qian, and
  Sheng]{chen_determining_2015}
Chen,~S.; Wang,~H.; Qian,~T.; Sheng,~P. Determining hydrodynamic boundary
  conditions from equilibrium fluctuations. \emph{Physical Review E}
  \textbf{2015}, \emph{92}, 043007\relax
\mciteBstWouldAddEndPuncttrue
\mciteSetBstMidEndSepPunct{\mcitedefaultmidpunct}
{\mcitedefaultendpunct}{\mcitedefaultseppunct}\relax
\EndOfBibitem
\bibitem[Omori \latin{et~al.}(2019)Omori, Inoue, Joly, Merabia, and
  Yamaguchi]{omori_full_2019}
Omori,~T.; Inoue,~N.; Joly,~L.; Merabia,~S.; Yamaguchi,~Y. Full
  characterization of the hydrodynamic boundary condition at the atomic scale
  using an oscillating channel: {Identification} of the viscoelastic
  interfacial friction and the hydrodynamic boundary position. \emph{Physical
  Review Fluids} \textbf{2019}, \emph{4}, 114201\relax
\mciteBstWouldAddEndPuncttrue
\mciteSetBstMidEndSepPunct{\mcitedefaultmidpunct}
{\mcitedefaultendpunct}{\mcitedefaultseppunct}\relax
\EndOfBibitem
\bibitem[Navier(1823)]{navier_memoire_1823}
Navier,~C.~L. \emph{M\'emoire de l'acad\'emie des sciences de l'institut de
  {France}}; Imprimerie royale, 1823; pp 389--440\relax
\mciteBstWouldAddEndPuncttrue
\mciteSetBstMidEndSepPunct{\mcitedefaultmidpunct}
{\mcitedefaultendpunct}{\mcitedefaultseppunct}\relax
\EndOfBibitem
\bibitem[Craig \latin{et~al.}(2001)Craig, Neto, and
  Williams]{craig_shear-dependent_2001}
Craig,~V. S.~J.; Neto,~C.; Williams,~D. R.~M. Shear-{Dependent} {Boundary}
  {Slip} in an {Aqueous} {Newtonian} {Liquid}. \emph{Physical Review Letters}
  \textbf{2001}, \emph{87}, 054504\relax
\mciteBstWouldAddEndPuncttrue
\mciteSetBstMidEndSepPunct{\mcitedefaultmidpunct}
{\mcitedefaultendpunct}{\mcitedefaultseppunct}\relax
\EndOfBibitem
\bibitem[Chennevi\`ere \latin{et~al.}(2016)Chennevi\`ere, Cousin, Bou\'e,
  Drockenmuller, Shull, L\'eger, and Restagno]{chenneviere_direct_2016}
Chennevi\`ere,~A.; Cousin,~F.; Bou\'e,~F.; Drockenmuller,~E.; Shull,~K.~R.;
  L\'eger,~L.; Restagno,~F. Direct {Molecular} {Evidence} of the {Origin} of
  {Slip} of {Polymer} {Melts} on {Grafted} {Brushes}. \emph{Macromolecules}
  \textbf{2016}, \emph{49}, 2348--2353\relax
\mciteBstWouldAddEndPuncttrue
\mciteSetBstMidEndSepPunct{\mcitedefaultmidpunct}
{\mcitedefaultendpunct}{\mcitedefaultseppunct}\relax
\EndOfBibitem
\bibitem[Mhetar and Archer(1998)Mhetar, and Archer]{mhetar_slip_1998}
Mhetar,~V.; Archer,~L.~A. Slip in entangled polymer solutions.
  \emph{Macromolecules} \textbf{1998}, \emph{31}, 6639--6649\relax
\mciteBstWouldAddEndPuncttrue
\mciteSetBstMidEndSepPunct{\mcitedefaultmidpunct}
{\mcitedefaultendpunct}{\mcitedefaultseppunct}\relax
\EndOfBibitem
\bibitem[Plucktaveesak \latin{et~al.}(1999)Plucktaveesak, Wang, and
  Halasa]{plucktaveesak_interfacial_1999}
Plucktaveesak,~N.; Wang,~S.-Q.; Halasa,~A. Interfacial {Flow} {Behavior} of
  {Highly} {Entangled} {Polybutadiene} {Solutions}. \emph{Macromolecules}
  \textbf{1999}, \emph{32}, 3045--3050\relax
\mciteBstWouldAddEndPuncttrue
\mciteSetBstMidEndSepPunct{\mcitedefaultmidpunct}
{\mcitedefaultendpunct}{\mcitedefaultseppunct}\relax
\EndOfBibitem
\bibitem[Sanchez-Reyes and Archer(2003)Sanchez-Reyes, and
  Archer]{sanchez-reyes_interfacial_2003}
Sanchez-Reyes,~J.; Archer,~L.~A. Interfacial {Slip} {Violations} in {Polymer}
  {Solutions}: {Role} of {Microscale} {Surface} {Roughness}. \emph{Langmuir}
  \textbf{2003}, \emph{19}, 3304--3312\relax
\mciteBstWouldAddEndPuncttrue
\mciteSetBstMidEndSepPunct{\mcitedefaultmidpunct}
{\mcitedefaultendpunct}{\mcitedefaultseppunct}\relax
\EndOfBibitem
\bibitem[Pearson and Petrie(1968)Pearson, and Petrie]{pearson_melt-flow_1968}
Pearson,~J.; Petrie,~C. \emph{Polymer systems: deformation and flow}, macmillan
  ed.; Wetton RE, Whorlow RH, 1968; pp 163--187\relax
\mciteBstWouldAddEndPuncttrue
\mciteSetBstMidEndSepPunct{\mcitedefaultmidpunct}
{\mcitedefaultendpunct}{\mcitedefaultseppunct}\relax
\EndOfBibitem
\bibitem[Hill \latin{et~al.}(1990)Hill, Hasegawa, and Denn]{hill_apparent_1990}
Hill,~D.~A.; Hasegawa,~T.; Denn,~M.~M. On the apparent relation between
  adhesive failure and melt fracture. \emph{Journal of Rheology} \textbf{1990},
  \emph{24}\relax
\mciteBstWouldAddEndPuncttrue
\mciteSetBstMidEndSepPunct{\mcitedefaultmidpunct}
{\mcitedefaultendpunct}{\mcitedefaultseppunct}\relax
\EndOfBibitem
\bibitem[Hatzikiriakos and Dealy(1991)Hatzikiriakos, and
  Dealy]{hatzikiriakos_wall_1991}
Hatzikiriakos,~S.~G.; Dealy,~J.~M. Wall slip of molten high density
  polyethylene. {I}. {Sliding} plate rheometer studies. \emph{Journal of
  Rheology} \textbf{1991}, \emph{35}, 497--523\relax
\mciteBstWouldAddEndPuncttrue
\mciteSetBstMidEndSepPunct{\mcitedefaultmidpunct}
{\mcitedefaultendpunct}{\mcitedefaultseppunct}\relax
\EndOfBibitem
\bibitem[De~Gennes(1979)]{de_gennes_ecoulements_1979}
De~Gennes,~P.-G. \'Ecoulements viscom\'etriques de polymères enchev\^etr\'es.
  \emph{C. R. Acad. Sc. Paris} \textbf{1979}, 219--220\relax
\mciteBstWouldAddEndPuncttrue
\mciteSetBstMidEndSepPunct{\mcitedefaultmidpunct}
{\mcitedefaultendpunct}{\mcitedefaultseppunct}\relax
\EndOfBibitem
\bibitem[Bäumchen \latin{et~al.}(2009)Bäumchen, Fetzer, and
  Jacobs]{baumchen_reduced_2009}
Bäumchen,~O.; Fetzer,~R.; Jacobs,~K. Reduced {Interfacial} {Entanglement}
  {Density} {Affects} the {Boundary} {Conditions} of {Polymer} {Flow}.
  \emph{Physical Review Letters} \textbf{2009}, \emph{103}, 247801\relax
\mciteBstWouldAddEndPuncttrue
\mciteSetBstMidEndSepPunct{\mcitedefaultmidpunct}
{\mcitedefaultendpunct}{\mcitedefaultseppunct}\relax
\EndOfBibitem
\bibitem[Housmans \latin{et~al.}(2014)Housmans, Sferrazza, and
  Napolitano]{housmans_kinetics_2014}
Housmans,~C.; Sferrazza,~M.; Napolitano,~S. Kinetics of {Irreversible} {Chain}
  {Adsorption}. \emph{Macromolecules} \textbf{2014}, \emph{47},
  3390--3393\relax
\mciteBstWouldAddEndPuncttrue
\mciteSetBstMidEndSepPunct{\mcitedefaultmidpunct}
{\mcitedefaultendpunct}{\mcitedefaultseppunct}\relax
\EndOfBibitem
\end{mcitethebibliography}


\providecommand{\latin}[1]{#1}
\makeatletter
\providecommand{\doi}
  {\begingroup\let\do\@makeother\dospecials
  \catcode`\{=1 \catcode`\}=2 \doi@aux}
\providecommand{\doi@aux}[1]{\endgroup\texttt{#1}}
\makeatother
\providecommand*\mcitethebibliography{\thebibliography}
\csname @ifundefined\endcsname{endmcitethebibliography}
  {\let\endmcitethebibliography\endthebibliography}{}

\end{document}